\documentclass[11pt]{amsart}

\usepackage{bbm}
\usepackage{bm}

\newcommand{\ov}[1]{\overline{#1}}

\DeclareMathOperator{\im}{Im}

\newcommand{\EMPTY}[1]{}

\newtheorem{Theorem}{Theorem}[section]

\newtheorem{Definition}[Theorem]{Definition}
\newtheorem{Example}[Theorem]{Example}
\newtheorem{Remark}[Theorem]{Remark}
\numberwithin{equation}{section}

\newcommand{\s}{\sigma}

\newcommand{\M}{{\mathcal M}}

\newcommand{\ty}{\infty}

\newcommand{\I}{\mathbbm{i}}
\newcommand{\D}{\mathrm{d}}

\newcommand{\eR}{\mathbb{R}}
\newcommand{\eN}{\mathbb{N}}
\newcommand{\Ze}{\mathbb{Z}}

\newcommand{\Ce}{\mathbb{C}}

\newcommand{\re}{\mathop{\mathrm{Re}}}

\newcommand{\po}{{\mathop{\mathcal P}}}

\newcommand{\res}{\operatorname{res}}

\newcommand{\ovb}[1]{\mkern 1.5mu\overline{\mkern-1.5mu#1\mkern-1.5mu}\mkern 1.5mu}
\newcommand{\unb}[1]{\mkern 1.5mu\underline{\mkern-1.5mu#1\mkern-1.5mu}\mkern 1.5mu}

\begin{document}

\title[Fractal tube formulas and a Minkowski measurability criterion]{Fractal tube formulas and a Minkowski measurability criterion for compact subsets of Euclidean spaces}

\author[M.~L.~Lapidus, G.~Radunovi\'c and D.~\v Zubrini\'c]{Michel~L.~Lapidus, Goran~Radunovi\'c and Darko~\v Zubrini\'c}

\thanks{The research of Michel L.~Lapidus was partially supported by the National Science
Foundation under grants ~DMS-0707524 and DMS-1107750, as well as by the Institut des Hautes \' Etudes Scientifiques (IH\' ES) where the first author was a visiting professor in the Spring of 2012 while part of this research was completed.}

\thanks{The research of Goran Radunovi\'c and Darko \v Zubrini\'c was supported in part by the Croatian Science Foundation under the project IP-2014-09-2285 and by the Franco-Croatian 
PHC-COGITO project.}

\begin{abstract}
We establish pointwise and distributional fractal tube formulas for a large class of compact subsets of Euclidean spaces of arbitrary dimensions.
These formulas are 
expressed as sums of residues of suitable meromorphic functions over the complex dimensions of the compact set under consideration (i.e., over the poles of its fractal zeta function).
Our results generalize to higher dimensions (and in a significant way) the corresponding ones previously obtained for fractal strings by the first author and van Frankenhuijsen.
They are illustrated by several examples and applied to yield a new Minkowski measurability criterion.
\end{abstract}

\maketitle

\section{Introduction}\label{intro0}
\setcounter{footnote}{0}

The development of the higher-dimensional theory of complex dimensions in \cite{fzf,larazu1,larazu2,memoir,cras2,mm,brezis,tabarz}
provides, among many other things, a new approach to the elusive notion of `fractality'.
To be more specific, for a set to be considered fractal, a commonly accepted proposal is that it should have a nontrivial fractal dimension, i.e., greater than its topological dimension; see, especially, \cite{Man}, where the Hausdorff dimension was used.
The problem is that none of the known fractal dimensions (Hausdorff, Minkowski, packing, etc.) gives a satisfactory character to all of the sets that we would like to call fractal.
Namely, the most famous counterexample is the devil's staircase; i.e., the Cantor function graph, for which all of the known fractal dimensions are equal to 1, its topological dimension.

The complex dimensions of bounded subsets generalize the notion of Minkowski dimension (also known as the box dimension).
More specifically, complex dimensions are defined as the multiset of (visible) poles (or more general singularities) of the associated fractal distance (or tube) zeta function and, under mild conditions, the Minkowski dimension of a bounded set is also its (real) complex dimension with maximal real part.
More importantly, the complex dimensions play a major role in determining the asymptotics of the volume of the $t$-neighborhoods of the given bounded set $A$ as $t\to0^+$ and therefore reflect its inner geometry.
This latter part of the theory of complex dimensions is announced in the present article, along with sketches of proofs of the main results and illustrations of the results in two examples.
The full proofs, and in the much greater generality of relative fractal drums, are provided in two articles \cite{cras2,mm} and can also be found in the research monograph \cite[Ch.\ 5]{fzf}.

The resulting `fractal tube formulas' can be interpreted pointwise or distributionally, as well as with or without an error term; see Theorems \ref{pointwise_formula_d} and \ref{dist_error_d}, along with Remark \ref{2.21/2}.
They enable us to express (modulo a possible error term) the volume $|A_t|$ of the $t$-neighborhood as a sum over the underlying complex dimensions $\omega$ of $A$ and the associated residues of the given fractal zeta function.
For example, in the case of simple poles, the corresponding exact fractal tube formula can be stated as follows (see Remark \ref{2.21/2}$(c)$):
\begin{equation}\label{1..1}
|A_t|=\sum_\omega c_\omega\frac{t^{N-\omega}}{N-\omega},
\end{equation}
where the sum runs over all of the complex dimensions of $A$ and $c_\omega$ is the residue of $\zeta_A$ (the distance zeta function of $A$, see Definition \ref{zeta_r}) evaluated at $\omega$.
In this manner, in Equation \eqref{1..1}, the oscillations that are intrinsic to fractal geometry are expressed in a natural way via the complex dimensions.
More specifically, the real parts (resp., the imaginary parts) of the underlying complex dimensions correspond to the amplitudes (resp., the frequencies) of the corresponding `geometric vibrations'.

We should stress that the fractal tube formulas obtained in this paper (see Theorems \ref{pointwise_formula_d}, \ref{dist_error_d} and Remark \ref{2.21/2}) extend to arbitrary dimensions $N\geq 1$ the corresponding ones obtained for fractal strings in \cite[\S 8.1]{lapidusfrank12} (i.e., for $N=1$).

Moreover, they also extend to very general fractal sets the corresponding tube formulas obtained for fractal sprays (a natural higher-dimensional generalization of fractal strings, \cite{LapPo2}) in \cite{lappe2,lappewi1}.\footnote{The former statement is briefly justified at the end of this paper, while both statements are explained in detail in \cite{cras2} and \cite[Ch.\ 5]{fzf}}.

As is expected, being a generalization of the Minkowski dimension, the complex dimensions are closely connected to the property of Minkowski measurability and this connection is, along with the aforementioned fractal tube formulas, the main announced result of the present paper; see Theorem \ref{criterion}.
Namely, under suitable hypotheses, we show that a bounded subset of $\eR^N$ is Minkowski measurable if and only if its only complex dimension with maximal real part $D$ is $D$ itself, its Minkowski (or box) dimension, and it is simple.

We note that the Minkowski measurability criterion obtained in Theorem \ref{criterion} extends to any dimension $N\geq 1$ the corresponding one obtained in \cite[\S 8.3]{lapidusfrank12} for fractal strings (i.e., when $N=1$).

The question of the Minkowski measurability of a given set $A$ has attracted considerable interest in the past.
Mandelbrot's suggestion in~\cite{Man2} to use the Minkowski content as a characteristic for the texture (or `lacunarity') of sets (see \cite{Man2} and \cite[\S X]{Man}) is one of the motivations for this question; more information on this subject can be found, e.g., in\cite{BedFi,Fr,gatzouras} and in~\cite[\S 12.1.3]{lapidusfrank12}.

More directly relevant to our present work, a lot of attention was devoted to the notion of Minkowski content in connection to the (modified) Weyl--Berry conjecture \cite{Lap1}.
This conjecture relates the spectral asymptotics of the Laplacian on a bounded open set and the Minkowski content of its boundary and it is resolved affirmatively in dimension one, i.e., for fractal strings in \cite{LaPo1, lapiduspom}.\footnote{For the original Weyl--Berry conjecture and its physical applications, see Berry's papers~\cite{Berr1,Berr2}.
Furthermore, early mathematical work on this conjecture and its applications can be found in~\cite{BroCar, Lap1, Lap2, Lap3, lapiduspom, LapPo2, FlVa}.
For a more extensive list of later work, see~\cite[\S 12.5]{lapidusfrank12}.}  
The characterization of the Minkowski measurability of bounded subsets of $\eR$ obtained in~\cite{lapiduspom} was a crucial part of proving this result.\footnote{A new proof of a part of this result was given in \cite{Fal2} and, more recently, in \cite{winter}.}
In particular, this work led to a useful reformulation of the Riemann hypothesis, as announced in \cite{LaMa2}, in terms of an inverse spectral problem for fractal strings; see~\cite{LapMa2}.
See also the formulation given in~\cite{Lap1} 
and which was proved for compact subsets of $\eR$ in 1993 by M.\ L.\ Lapidus and C.\ Pomerance in~\cite{lapiduspom} (but disproved in higher dimensions, at least in its original form, in \cite{FlVa,LapPo2}; see, however, the corresponding conjectures made in \cite{Lap3} for self-similar drums and on which our present work may eventually help us shine new light).

In closing  this introduction, we mention that works involving the notion of Minkowski measurability (or of Minkowski content) include \cite{Tri1,Carfi,lapidushe,rae,gatzouras,BroCar,Fal2,FlVa,Fr,Kom,Lap1,Lap2,Lap3,lapz,lap8,LaMa2,LapMa2,lappe2,lappewi1,LaPeWi2,lapiduspom,LaPo1,LapPo2,fzf,larazu1,larazu2,memoir,cras2,mm,brezis,tabarz,lapidusfrank12,Man2,winter,Wi}, along with the many relevant references therein.
References on tube formulas in the `smooth setting' and in the `fractal setting' are provided at the beginning of Section \ref{sekcija_3}.

\section{Preliminaries}\label{intro}
We begin by stating some definitions and results from the papers \cite{larazu1,larazu2} and the research monograph~\cite{fzf} that will be needed in this article, as well as by recalling some well-known notions.
Given a bounded subset $A$ of $\eR^N$ (always assumed to be nonempty in this paper), we denote its {\em $\delta$-neighborhood} by $A_\delta:=\{x\in\eR^N:d(x,A)<\delta\}$.
Here, $d(x,A):=\inf\{|x-y|:y\in A\}$ is the Euclidean distance between the point $x$ and the set $A$.\footnote{Without loss of generality, we may replace $A$ by its closure, $\ov{A}$, and hence assume from now on that $A$ is compact.}
Furthermore, for a compact subset $A$ of $\eR^N$ and $r\geq 0$, we define its {\em upper $r$-dimensional Minkowski content},
\begin{equation}\label{2..1/2}
{\ovb{\mathcal{M}}}^{r}(A)=\limsup_{t\to0^+}\frac{|A_t|}{t^{N-r}},
\end{equation} and its {\em upper box dimension},
$$
\ovb{\dim}_BA=\inf\{r\geq 0:{\ovb{\mathcal{M}}}^{r}(A)=0\}=\sup\{r\geq 0:\ovb{\mathcal{M}}^{r}(A)=+\ty\}.
$$
The value $\unb{\mathcal{M}}^{r}(A)$ of the {\em lower $r$-dimensional Min\-kow\-ski content} of $A$ is defined analogously as ${\ovb{\mathcal{M}}}^{r}(A)$, except for a lower instead of an upper limit; and similarly for the {\em lower box dimension} $\unb{\dim}_BA$.

If $\unb{\dim}_BA=\ovb{\dim}_BA$, this common value is called the {\em Minkowski} (or {\em box}) {\em dimension} of $A$ and denoted by $\dim_BA$.
If $0<\unb{\mathcal{M}}^D(A)(\le)\ovb{\mathcal{M}}^{D}(A)<\ty$, for some $D\geq 0$, the set $A$ is said to be {\em Minkowski nondegenerate}.
It then follows that $\dim_BA$ exists and is equal to $D$.
Moreover, if $\mathcal{M}^D(A)$ exists and is different from $0$ and $+\ty$ (in which case $\dim_BA$ exists and then necessarily, $D=\dim_BA$), the set $A$ is said to be {\em Minkowski 
measurable}.

We will now introduce the notions of distance and tube zeta functions of compact sets and state their basic properties.
These definitions have enabled us in~\cite{larazu1,larazu2,memoir,cras2,mm,fzf,brezis} to develop a higher-dimensional extension of the theory of complex dimensions of fractal strings (\cite{lapidusfrank12}), valid for arbitrary compact sets.

\smallskip

\begin{Definition}[Fractal zeta functions,~{\cite{larazu1}}]\label{zeta_r} {\rm Let $A$ be a compact subset of $\eR^N$ and fix $\delta>0$.
We define the {\em distance zeta function $\zeta_A$} of $A$ and the {\em tube zeta function $\widetilde{\zeta}_A$} of $A$ by the following Lebesgue integrals, respectively, for some $\delta>0$ and for all $s\in\Ce$ with $\re s$ sufficiently large$:$
\begin{equation}\label{401/2}
\zeta_A(s;\delta):=\int_{A_\delta} d(x,A)^{s-N}\D x\quad\textrm{and}\quad\widetilde{\zeta}_A(s;\delta):=\int_0^{\delta}t^{s-N-1}|A_t|\,\D t.
\end{equation}
}
\end{Definition}

\smallskip

It is not difficult to show that the distance and tube zeta functions of a compact subset $A$ of $\eR^N$ satisfy the following functional equation, which is valid on any connected open set $U\subseteq\Ce$ containing the vertical line $\{\re s=\ov{\dim}_BA\}$ and to which any (and hence, both) of the two zeta functions has a meromorphic continuation (see~\cite{larazu1} or \cite[\S2.2]{fzf})$:$
\begin{equation}\label{equ_tilde}
\zeta_A(s;\delta)=\delta^{s-N}|A_\delta|+(N-s)\widetilde\zeta_A(s;\delta).
\end{equation}

\smallskip

Furthermore, in the above definition (see Equation \eqref{401/2}), the dependence of the zeta functions on the parameter $\delta>0$ is inessential, from the point of view of the theory of complex dimensions $($see Definition \ref{com_dim} below$)$. 
Indeed, it is shown in \cite{fzf} that the difference of two distance (or tube) zeta functions of the same compact set $A$, and corresponding to any two different values of the parameter $\delta$, is an entire function. 

\smallskip

Let us briefly summarize the main properties of the distance and tube zeta functions (see~\cite{larazu1} or \cite[Ch.~2]{fzf}):

\smallskip

If $A$ is a compact subset of $\eR^N$, then the tube zeta function $\widetilde{\zeta}_A(\,\cdot\,;\delta)$ is holomorphic in the half-plane $\{\re s>\overline{\dim}_BA\}$ and $\ov{\dim}_BA$ coincides with the abscissa of (absolute) convergence of $\widetilde{\zeta}_A(\,\cdot\,;\delta)$.

 Furthermore, if the box $($or Minkowski$)$ dimension $D:=\dim_BA$ exists and $\unb{\M}^D(A)>0$, then $\widetilde{\zeta}_A(s;\delta)\to+\ty$ as $s\in\eR$
converges to $D$ from the right.
The above statements are also true if we replace  $\widetilde{\zeta}_A$ by ${\zeta}_A$ and in the preceding sentence assume, in addition, that $D<N$. Finally, we have the {\em scaling property}; that is, if for $\lambda>0$, we let $\lambda A:=\{\lambda x:x\in A\}$, then $\zeta_{\lambda A}(s;\lambda\delta)=\lambda^s\zeta_A(s;\delta)$ and $\widetilde{\zeta}_{\lambda A}(s;\lambda\delta)=\lambda^s\widetilde{\zeta}_A(s;\delta)$.

\smallskip

If $A$ is a Minkowski nondegenerate subset of $\eR^N$ $($so that $D:=\dim_BA$ exists$)$, 
and for some $\delta>0$ there exists a meromorphic extension of $\widetilde\zeta_A(\,\cdot\,;\delta)$ to a connected open neighborhood of $D$, then  $D$ is a simple  pole of $\widetilde{\zeta}_A(\,\cdot\,;\delta)$
and the residue $\res(\widetilde{\zeta}_A(\,\cdot\,;\delta),D)$ is independent of $\delta$.
Furthermore, we have
$$
\unb{\M}^D(A)\le\res(\widetilde\zeta_A(\,\cdot\,;\delta), D)\le {\ovb{\M}}^{D}(A).
$$
In particular, if $A$ is Minkowski measurable, then 
$$
\res(\widetilde\zeta_A(\,\cdot\,;\delta), D)=\M^D(A).
$$
If, additionally, $D<N$, the analogous statement and conclusion is true for the distance zeta function ${\zeta}_A$ and we have
$$
(N-D)\unb{\M}^D(A)\le\res(\zeta_A(\,\cdot\,;\delta),D)\le(N-D){\ovb{\M}}^{D}(A).
$$
Moreover, if $A$ is Minkowski measurable, then 
$$
\res(\zeta_A(\,\cdot\,;\delta), D)=(N-D)\M^D(A).
$$

\smallskip

We refer to \cite{larazu2} as well as to \cite[\S 2.2]{fzf} for sufficient conditions guaranteeing the existence of a meromorphic continuation for $\zeta_A$ (or, equivalently, provided $\ov{\dim}_BA<N$, $\widetilde{\zeta}_A$) on a suitable domain.

\smallskip

Let us now introduce some additional definitions, which are adapted from~\cite{lapidusfrank12} to the present, much more general, context of compact subsets of an arbitrary Euclidean space, $\eR^N$ (with $N\geq 1$):

\smallskip

The {\em screen} $\bm{S}$ is the graph of a bounded, real-valued, Lipschitz continuous function $S(\tau)$, with the horizontal and vertical axes interchanged:
$$\bm{S}:=\{S(\tau)+\I \tau\,:\,\tau\in\eR\}.$$
The Lipschitz constant is denoted by $\|S\|_{\mathrm{Lip}}$.
Furthermore, we let $$\sup S:=\sup_{\tau\in\eR}S(\tau)\in\eR.$$
For a compact subset $A$ of $\eR^N$, we always assume that the screen $\bm{S}$ lies to the left of the {\em critical line} $\{\re s=\ov{D}\}$, i.e., that $\sup S\leq\ov{D}$.
Moreover, the {\em window} $\bm{W}$ is the closed subset of $\Ce$ defined by $$\bm{W}:=\{s\in\Ce:\re s\geq S(\im s)\}.$$
The set $A$ is said to be {\em admissible} if its tube (or distance) zeta function can be meromorphically extended to an open connected neighborhood of some window $\bm{W}$ and does not have any pole located on the corresponding screen $\bm{S}$.

\smallskip

\begin{Definition}[$d$-languid set; adapted from~{\cite[Def.~5.2]{lapidusfrank12}}]\label{languid}
{\rm An admissible compact subset $A$ of $\eR^N$ is said to be {\em $d$-languid} if there exists $\delta>0$ such that ${\zeta}_A(s;\delta)$ satisfies the following growth conditions: There exist real constants $\kappa_d$ and $C>0$ and a two-sided sequence $(T_n)_{n\in\Ze}$ of real numbers such that $T_{-n}<0<T_n$ for $n\geq 1$,
$\lim_{n\to\ty}T_n=+\ty$ and $\lim_{n\to\ty}T_{-n}=-\ty
$,
satisfying the following two hypotheses, {\bf L1} and {\bf L2}:
}

\smallskip

{\bf L1} There exists $c>N$ such that
$|{\zeta}_{A}(\s+\I T_n;\delta)|\leq C(|T_n|+1)^{\kappa_d}$, for all $n\in\Ze$ and all $\s\in (S(T_n),c)$.

\smallskip

{\bf L2} For all $\tau\in\eR$, with $|\tau|\geq 1$, we have that
$|{\zeta}_{A}(S(\tau)+\I \tau;\delta)|\leq C|\tau|^{\kappa_d}$.
\end{Definition}

\smallskip

\begin{Definition}[Strongly $d$-languid set; adapted from~{\cite[Def.~5.3]{lapidusfrank12}}]\label{str_languid}
{\rm A compact subset $A$ of $\eR^N$ is said to be {\em strongly $d$-languid} if for some $\delta>0$, ${\zeta}_A(s;\delta)$ satisfies {\bf L1} with $S(\tau)\equiv -\ty$ in condition {\bf L1}; i.e., for every $\sigma<c$ and, additionally, there exists a sequence of screens $S_m(\tau)\colon \tau\mapsto S_m(\tau)+\I \tau$ for $m\geq 1$, $\tau\in\eR$ with $\sup S_m\to -\ty$ as $m\to\ty$ and with a uniform Lipschitz bound, $\sup_{m\geq 1}\|S_m\|_{\mathrm{Lip}}<\ty$, such that

\medskip

{\bf L2'} There exist $B,C>0$ such that
$$
|{\zeta}_{A}(S_m(\tau)+\I \tau;\delta)|\leq CB^{|S_m(\tau)|}(|\tau|+1)^{\kappa_d},
$$ for all $\tau\in\eR$ and $m\geq 1$.}
\end{Definition}

\medskip

\begin{Definition}[Complex dimensions,~{\cite{fzf,larazu2}}]\label{com_dim}
{\rm Let $A$ be an admissible compact subset of $\eR^N$. 
Then, the set of {\em visible complex dimensions of $A$ $($with respect to $U)$} is defined as
$$
\po({\zeta}_A(\,\cdot\,;\delta),U):=\{\omega\in U:\omega\textrm{ is a pole of }{\zeta}_A(\,\cdot\,;\delta)\}.
$$
If $U=\Ce$, we say that $\po({\zeta}_A(\,\cdot\,;\delta),\Ce)$ is the set of {\em complex dimensions} of $A$.\footnote{Clearly, $\po({\zeta}_A(\,\cdot\,;\delta),U)$ is a discrete subset of $\Ce$ and is independent of $\delta$; hence, so is $\po({\zeta}_A(\,\cdot\,;\delta),\Ce)$.
Therefore, we will often write $\po({\zeta}_A,U)$ or $\po({\zeta}_A,\Ce)$ instead.}
} 
\end{Definition}

\section{Pointwise and distributional tube formulas and a criterion for Minkowski measurability}\label{sekcija_3}

In this section, we state and sketch the proof of our main results, the pointwise and distributional tube formulas, valid for a large class of compact subsets of $\eR^N$ (see Theorem \ref{pointwise_formula_d} and \ref{dist_error_d} below), along with an associated Minkowski measurability criterion (see Theorem \ref{criterion}).
These results extend to higher dimensions the corresponding tube formulas and Minkowski measurability criterion obtained for fractal strings in \cite{lapidusfrank12}, \S8.1 and \S8.3, respectively.
We point out that the detailed proofs of our main results (stated in a much more general form and within the broader context of relative fractal drums) can be found in the long papers corresponding to this article, \cite{mm,cras2}.
Moreover, we note that in light of the functional equation \eqref{equ_tilde}, Theorems \ref{pointwise_formula_d}, \ref{dist_error_d} and \ref{criterion} have an obvious analog for tube (instead of distance) zeta functions.
Also, the exact tube formula stated in Theorem \ref{pointwise_formula_d} has a counterpart with error term (much as in Theorem \ref{dist_error_d}); see Remark \ref{2.21/2}$(b)$.
Finally, we refer to \cite{mm,cras2} and \cite[\S13.1]{lapidusfrank12} for many additional references on tube formulas (or related formulas) in various settings, including, \cite{Kow,Bla,Fed1,KlRot,minkow,Ol1,Ol2,Za5,Wi,DeKoOzUr,Gra,HuLaWe,Schn,Zah,lappe2,lappewi1,LaPeWi2,fzf,mm,cras2,WiZa,Wey3,Stein,tabarz}.

\smallskip

The key observation in deriving Theorem \ref{pointwise_formula_d} and \ref{dist_error_d} below is the fact that the tube zeta function of a compact set $A$ in $\eR^N$ is equal to the Mellin transform of its modified tube function 
\begin{equation}
f(t):=\chi_{(0,\delta)}(t)t^{-N}|A_t|,
\end{equation} 
where $\chi_E$ denotes the characteristic function of the set $E$.
More precisely, one has that 
\begin{equation}
\widetilde{\zeta}_A(s;\delta)=\{\mathfrak{M} f\}(s):=\int_0^{+\ty}t^{s-1}f(t)\,\D t,
\end{equation}
where $\mathfrak{M}$ denotes the Mellin transform.
One then applies the Mellin inversion theorem (see, e.g.,~\cite[Thm.~28]{titch}) to deduce that
\begin{equation}
|A_t|=\frac{1}{2\pi\I}\int_{c-\I\ty}^{c+\I\ty}t^{N-s}\widetilde{\zeta}_A(s;\delta)\,\D s,
\end{equation}
for all $t\in(0,\delta)$,
where $c>\ovb{\dim}_BA$ is arbitrary.
One then proceeds in a similar manner, much as in~\cite[Ch.~5]{lapidusfrank12} for the case of fractal strings.
More precisely, one works with a $k$-th primitive function of $t\mapsto|A_t|$ in order to be able to represent the above integral as a sum over the complex dimensions contained in the window $\bm{W}$.
Here, $k\in\eN$ is taken large enough to ensure the pointwise convergence of this sum.
From this result, one then derives (by distributional differentiation) the distributional tube formula for every value of $k$ (even for $k\in\Ze$), and, in particular, for $k=0$.
In this way, we obtain the fractal tube formulas expressed in terms of the tube zeta function and then use the functional equation~\eqref{equ_tilde} in order to translate them in terms of the distance zeta function.

\smallskip

\begin{Theorem}[Pointwise tube formula]\label{pointwise_formula_d}
Let $A$ be a compact subset of $\eR^N$ such that $\ovb{\dim}_BA<N$.
Furthermore, assume that there exists  a constant $\lambda> 0$ such that $\lambda A$ is strongly $d$-languid for some $\delta>0$ and $\kappa_d<1$ $($with $\kappa_d\in\eR$$)$.
Then, for every $t\in(0,\lambda^{-1}\min\{1,\delta,B^{-1}\})$, the following exact pointwise tube formula is valid $($where $B$ is the constant appearing in {\bf L2'} of Definition \ref{str_languid}  above$)$$:$\footnote{Here and in Theorem \ref{dist_error_d}, we write $\zeta_A(s)$ instead of $\zeta_A(s;\delta)$ since the residues in the formula do not depend on the parameter $\delta$.}
\begin{equation}\label{point_form_d}
|A_t|=\sum_{\omega\in\po({\zeta}_A,\Ce)}\res\left(\frac{t^{N-s}}{N-s}{\zeta}_A(s),\omega\right).
\end{equation}
\end{Theorem}

\smallskip

In the case when $\kappa_d\in\eR$, we usually only have a distributional tube formula.
Furthermore, if $A$ is only $d$-languid, we will have a distributional error term, with information about its asymptotic order given in the sense of \cite[\S5.4]{lapidusfrank12}. 
Namely, the distribution $\mathcal R\in\mathcal D'(0,\delta)$ is said to be of {\em asymptotic order at most} $t^{\alpha}$ $($resp., {\em less than} $t^{\alpha}$$)$ as $t\to 0^+$ if when applied to a test function $\varphi\in\mathcal{D}(0,\delta)$,\footnote{
Here, $\mathcal{D}(0,\delta):=C_c^{\ty}(0,\delta)$ is the standard space of infinitely differentiable test functions with compact support.} we have that
$
\langle\mathcal R,\varphi_a\rangle=O(a^{\alpha})
$
(resp.,
$
\langle\mathcal R,\varphi_a\rangle=o(a^{\alpha}))
$,
as $a\to 0^+$,
where $\varphi_a(t):=a^{-1}\varphi(t/a)$ (and the implicit constants may depend on $\varphi$).
We then write that $\mathcal R(t)=O(t^{\alpha})$ $($resp., $\mathcal R(t)=o(t^{\alpha}))$ as $t\to 0^+$.

\smallskip

\begin{Theorem}[Distributional tube formula]\label{dist_error_d}
Let $A$ be a $d$-languid compact subset of $\eR^N$, for some $\delta>0$ and $\kappa_d\in\eR$.
Furthermore, assume that $\ovb{\dim}_BA<N$ and denote by $\mathcal V(t)$ the $($regular$)$ distribution generated by the locally integrable function $t\mapsto|A_t|$.
Then, we have the following distributional equality$:$
\begin{equation}\label{dist_form_error_d}
\mathcal V(t)=\sum_{\omega\in\po({\zeta}_A,\bm{W})}\res\left(\frac{t^{N-s}}{N-s}{\zeta}_A(s),\omega\right)+\mathcal R(t).
\end{equation}
More precisely, the action of $\mathcal V(t)$ on a test function $\varphi\in\mathcal{D}(0,+\ty)$ is given by
\begin{equation}\label{error_action_}
\begin{aligned}
\big\langle\mathcal V,\varphi\big\rangle&=\sum_{\omega\in\po({\zeta}_A,\bm{W})}\res\left(\frac{\{\mathfrak{M}\varphi\}(N-s+1)}{N-s}{\zeta}_A(s),\omega\right)+\big\langle\mathcal R,\varphi\big\rangle.
\end{aligned}
\end{equation}
In Equation \eqref{dist_form_error_d}, the distributional error term $\mathcal R(t)$ is $O(t^{N-\sup S})$ as $t\to 0^+$.
Moreover, if $S(\tau) < \sup S$ for all $\tau\in\eR$, then $\mathcal R(t)$ is $o(t^{N-\sup S})$ as $t\to 0^+$.
If, in addition, $\lambda A$ is strongly $d$-languid for some $\lambda>0$, then, for all test functions in $\mathcal{D}\big(0,\lambda^{-1}\min\{1,\delta,B^{-1}\}\big)$, we have that $\mathcal{R}\equiv 0$ and $\bm{W}=\Ce;$ hence, we obtain an {\em exact} tube formula in that case. 
\end{Theorem}

\begin{Remark}\label{2.21/2}{\rm $(a)$\ Theorems \ref{pointwise_formula_d} and \ref{dist_error_d} have natural counterparts for the $k$-th primitive of $|A_t|$ or of $\mathcal{V}(t)$ (where $k\in\eN_0:=\eN\cup\{0\}$ for the pointwise formula, and $k\in\Ze$ for the distributional formula, with negative values of $k$ denoting the $|k|$-th distributional derivative of $\mathcal{V}(t)$).
In the analog of Theorem \ref{pointwise_formula_d}, the larger $k$, the weaker the assumptions on the growth of $\zeta_A$.
Furthermore, for the counterpart of Theorem \ref{dist_error_d}, the special case when $k=-1$, corresponding to the `geometric density of states' $\D\mathcal{V}/\D t$, viewed as a positive measure, is conceptually the most important one.

$(b)$\ If in Theorem \ref{pointwise_formula_d}, we assume the weaker hypothesis that $\zeta_A$ is $d$-languid relative to a window $\bm{W}$, then (under suitable assumptions on $\kappa_d$), we obtain a pointwise fractal tube formula with error term $R(t)$.
Furthermore, the sum in the analog of \eqref{point_form_d} is then taken over the set of visible complex dimensions, $\po(\zeta_A,\bm{W})$.
Moreover, the error term can be estimated much as in Theorem \ref{dist_error_d}, but now pointwise instead of distributionally.

$(c)$\ The special case of Theorem \ref{pointwise_formula_d} when all of the complex dimensions are simple yields the pointwise fractal tube formula stated in Equation \eqref{1..1} of the introduction (i.e., \S\ref{intro0}):
\begin{equation}\label{2.61/2}
|A_t|=\sum_{\omega\in\po(\zeta_A,\Ce)}c_\omega \frac{t^{N-\omega}}{N-\omega},
\end{equation}
where $c_\omega=\res(\zeta_A,\omega)$ for each $\omega\in\po(\zeta_A,\Ce)$.
Naturally, depending on the hypotheses, Equation \eqref{2.61/2} has a pointwise analog with error term, as well as a distributional analog with or without error term.
}
\end{Remark}

\smallskip

One of the applications of the above results is a Minkowski measurability criterion for a compact $d$-languid subset of $\eR^N$ (see Theorem~\ref{criterion} below), which generalizes~\cite[Thm.~8.15]{lapidusfrank12} to higher dimensions.
In the proof of Theorem~\ref{criterion}, one direction is a consequence of the distributional tube formula (Theorem~\ref{dist_error_d} above) and the uniqueness theorem for almost periodic distributions (see~\cite[\S VI.9.6, p.\ 208]{Schw}).
The other direction follows from a generalization of the classic Wiener--Ikehara Tauberian theorem (see \cite{Kor}) and for this direction of Theorem \ref{criterion} to hold, the assumptions on the distance zeta function $\zeta_A$ can be considerably weakened (see \cite{mm}).

\smallskip

\begin{Theorem}[Minkowski measurability criterion]\label{criterion}
Let $A$ be a compact subset of $\eR^N$ such that $D:=\dim_BA$ exists and $D<N$.
Furthermore, assume that $A$ is $d$-languid for a screen passing between the critical line $\{\re s=D\}$ and all the complex dimensions of $A$ with real part strictly less than $D$.
Then, the following statements are equivalent$:$

\medskip

$(a)$ $A$ is Minkowski measurable.

\medskip

$(b)$ $D$ is the only pole of ${\zeta}_A$ located on the critical line $\{\re s=D\}$, and it is simple.
\end{Theorem}

\medskip

There exist $d$-languid compact sets (and even fractal strings, see \cite[Exple.~5.32]{lapidusfrank12}) which do not satisfy the hypothesis of Theorem~\ref{criterion} concerning the screen. 
We point out that Theorems~\ref{pointwise_formula_d} and \ref{dist_error_d} can be applied to obtain tube formulas for a variety of well-known fractal sets, as is illustrated by the following examples.
Furthermore, Example~\ref{ex2} below shows how our results can be applied to derive the tube formula of a self-similar fractal set in $\eR^3$.
We further note that fractal tube formulas can also be obtained for examples of higher-dimensional fractal sets that are not self-similar, such as ``fractal nests'' and ``geometric chirps'', as well as for a version of the Cantor graph or devil's staircase (see \cite{fzf,memoir} for the definitions of these notions).
In such examples, we will generally obtain a distributional (or pointwise) tube formula with an error term; see \cite{cras2} and \cite[Ch.\ 5]{fzf} for details.



\smallskip

\begin{Example}\label{ex1}
{\rm Let $A$ be the Sierpi\'nski gasket in $\eR^2$, constructed in the usual way inside the unit triangle.
Then, for $\delta>1/4\sqrt{3}$, the distance zeta function $\zeta_A$ is given for all $s\in\Ce$ by
\begin{equation}\nonumber
\zeta_A(s;\delta)=\frac{6(\sqrt3)^{1-s}2^{-s}}{s(s-1)(2^s-3)}+2\pi\frac{\delta^s}s+3\frac{\delta^{s-1}}{s-1},
\end{equation}
which is meromorphic on the whole complex plane (see~\cite[\S 4.2]{memoir} or \cite[\S3.2]{fzf}). In particular, $\po({\zeta}_A,\Ce)=\{0\}\cup\big(\log_23+\frac{2\pi}{\log 2}\I\Ze\big)$, each complex dimension is simple, and by letting $\omega_k:=\log_23+\mathbf{p}k\I$ (so that $\omega_0=\log_32=\dim_BA$) and $\mathbf{p}:=2\pi/\log 2$, we have that
$$
\res(\zeta_A(\,\cdot\,;\delta),\omega_k)=\frac{6(\sqrt3)^{1-\omega_k}}{{4^{\omega_k}(\log2)\omega_k(\omega_k-1)}}
$$
(for all $k\in\Ze$) and
$
\res(\zeta_A(\,\cdot\,;\delta),0)=3\sqrt{3}+2\pi.
$
One can easily check, by using the scaling property of the distance zeta function, that $\lambda A$ is strongly $d$-languid, for any $\lambda \geq 2\sqrt{3}$ and with $\kappa_d:=-1$.
Hence, we can apply Theorem~\ref{pointwise_formula_d} in order to obtain the following exact pointwise tube formula, valid for all $t\in(0,1/2\sqrt{3})$, and which coincides with the one obtained in~\cite{lappe2,lappewi1,LaPeWi2} and also, in~\cite{DeKoOzUr}:
\begin{equation}\label{3.71/2}
\begin{aligned}
|A_t|&=\sum_{\omega\in\po({\zeta}_A,\Ce)}\res\left(\frac{t^{2-s}}{2-s}{\zeta}_A(s;\delta),\omega\right)\\
&=\frac{6\sqrt{3}\,t^{2-\log_23}}{\log 2}\sum_{k=-\ty}^{\ty}\frac{(4\sqrt{3})^{-\omega_k}t^{-\mathbf{p}k\I}}{(2-\omega_k)(\omega_k-1)\omega_k}+\left(\frac{3\sqrt{3}}{2}+\pi\right)t^2.
\end{aligned}
\end{equation}

By Theorem \ref{criterion} (and in accord with \cite{lappe2,lappewi1,LaPeWi2}), it follows that the Sierpi\'nski gasket is not Minkowski measurable.
Indeed, $D:=\dim_BA=\log_32$ is simple and $A$ has nonreal complex dimensions with real part $D$.
Note that the fact that $A$ is not Minkowski measurable also follows directly from \eqref{3.71/2} and the definition of Minkowski measurablity; see \eqref{2..1/2} and the discussion following it.}
\end{Example}

\smallskip

\begin{Example}\label{ex2}
{\rm Let $A$ be the three-dimensional analog of the Sierpi\'nski carpet.
More precisely, we construct $A$ by dividing the closed unit cube of $\eR^3$ into $27$ congruent cubes and remove the open middle cube, then we iterate this step with each of the $26$ remaining smaller closed cubes; and so on, ad infinitum.
By choosing $\delta>1/6$, we deduce that $\zeta_A$ is meromorphic on $\Ce$ and given for all $s\in\Ce$ by (see~\cite{fzf} or \cite{tabarz})
$$
\zeta_A(s;\delta)=\frac{48\cdot 2^{-s}}{s(s-1)(s-2)(3^s-26)}+\frac{4\pi\delta^s}{s}+\frac{6\pi\delta^{s-1}}{s-1}+\frac{6\delta^{s-2}}{s-2}.
$$
In particular, $\po({\zeta}_A,\Ce)=\{0,1,2\}\cup\big(\log_326+\mathbf{p}\I\Ze\big)$, where $\mathbf{p}:=2\pi/\log 3$; furthermore, each complex dimension of $A$ is simple.
Moreover, we have that $$\res(\zeta_A(\,\cdot\,;\delta),0)=4\pi-\frac{24}{25},\quad \res(\zeta_A(\,\cdot\,;\delta),1)=6\pi+\frac{24}{23},$$ $$\res(\zeta_A(\,\cdot\,;\delta),2)=\frac{96}{17}$$ and, by letting $\omega_k:=\log_326+\mathbf{p}k\I$ (for all $k\in\Ze$), $$\res(\zeta_A(\,\cdot\,;\delta),\omega_k)=\frac{24}{13\cdot 2^{\omega_k}\omega_k(\omega_k-1)(\omega_k-2)\log 3}.$$
One easily checks that the hypotheses of Theorem \ref{pointwise_formula_d} are satisfied with $\kappa_d:=-1$, and thus we obtain the following exact pointwise tube formula, valid for all $t\in(0,1/{2})$:
\begin{equation}\label{carp_tube}
\begin{aligned}
|A_t|&=\frac{24\,t^{3-\log_326}}{13\log 3}\sum_{k=-\ty}^{\ty}\frac{2^{-\omega_k}t^{-\mathbf{p}k\I}}{(3-\omega_k)(\omega_k-1)(\omega_k-2)\omega_k}\\
&\phantom{=}+\left(6-\frac{6}{17}\right)t+\left(3\pi+\frac{12}{23}\right)t^2+\left(\frac{4\pi}{3}-\frac{8}{25}\right)t^3.
\end{aligned}
\end{equation}
In particular, we conclude that $\dim_BA=\log_326$ and, by Theorem \ref{criterion}, that the three-dimensional Sierpi\'nski carpet is not Minkowski measurable (as expected).
Again, this conclusion follows from either the Minkowski measurability criterion provided in Theorem \ref{criterion} or directly from the definitions and the above fractal tube formula obtained in \eqref{carp_tube}.
}
\end{Example}

\smallskip

We conclude this paper by pointing out that, in a precise way, the above results generalize the corresponding ones obtained for fractal strings in \cite[\S8.1 \& \S8.3]{lapidusfrank12}.
Namely, this can be seen from the fact that for the geometric zeta function $\zeta_{\mathcal{L}}$ of a nontrivial fractal string $\mathcal{L}=(l_j)_{j\geq 1}$ and the distance zeta function of the set 
$$
A_{\mathcal{L}}:=\bigg\{a_k:=\sum_{j\geq k}l_j:k\geq 1\bigg\},
$$ we have that
$$
\zeta_{A_{\mathcal{L}}}(s;\delta)=\frac{2^{1-s}}{s}{\zeta_{\mathcal{L}}(s)}+\frac{2\delta^s}{s},
$$
where $\delta>l_1/2$, and this identity holds on any subdomain $U$ of $\Ce$ containing the critical line $\{\re s=\ov{\dim}_BA_{\mathcal{L}}\}$ and to which any of the two zeta functions has a meromorphic continuation; see~\cite[\S2.1]{fzf}.
Hence, if $U\subseteq\Ce\setminus\{0\}$, then $\zeta_{\mathcal{L}}$ and $\zeta_{A_{\mathcal{L}}}$ have the same visible complex dimensions in $U$, with the same multiplicities; furthermore, their corresponding residues (or, more generally, principal parts) are related in a straightforward manner.

\address{Michel L.\ Lapidus,\\ Department of Mathematics, University of California, Riverside,\\ California 92521-0135, USA\\
\email{lapidus@math.ucr.edu}}

\address{Goran Radunovi\'c,\\Department of Applied Mathematics, Faculty of Electrical Engineering and Computing, University of Zagreb\\ Unska 3, 10000 Zagreb, Croatia\\
\email{goran.radunovic@fer.hr}
}

\address{Darko \v Zubrini\'c,\\Department of Applied Mathematics, Faculty of Electrical Engineering and Computing, University of Zagreb\\ Unska 3, 10000 Zagreb, Croatia\\
\email{darko.zubrinic@fer.hr}\\
\end{document}